\documentclass{article}

\usepackage[final]{neurips_2020_ml4ps}

\usepackage[utf8]{inputenc} 
\usepackage[T1]{fontenc}    
\usepackage{hyperref}       
\usepackage{url}            
\usepackage{booktabs}       
\usepackage{amsfonts}       
\usepackage{nicefrac}       
\usepackage{microtype}      
\usepackage{mathtools}
\usepackage{graphicx}
\usepackage{amsmath}
\usepackage{geometry}
\usepackage{cleveref}
\usepackage{bm}
\usepackage{xcolor}
\usepackage{physics}        

\newcommand{\revised}[1]{#1}

\newcommand{\GRAPPA}{%
Gravitation Astroparticle Physics Amsterdam (GRAPPA)\\
Institute for Theoretical Physics (ITFA)\\
University of Amsterdam, Science Park 904, 1098 XH Amsterdam, The Netherlands}

\newcommand{\SISSA}{%
SISSA (Scuola Internazionale Superiore di Studi Avanzati)\\
via Bonomea 265, I-34136 Trieste, Italy.}

\title{Targeted Likelihood-Free Inference of Dark Matter Substructure in Strongly-Lensed Galaxies}

\author{%
  Adam Coogan\textsuperscript{1} \\
  \texttt{a.m.coogan@uva.nl} \\
  \And
  Konstantin Karchev\textsuperscript{1, 2} \\
  \texttt{kkarchev@sissa.it}
  \And
  Christoph Weniger\textsuperscript{1} \\
  \texttt{c.weniger@uva.nl} \\\\
  \textsuperscript{1}\GRAPPA \\\\
  \textsuperscript{2}\SISSA
}

\newcommand{\varmatrix}[1]{\ensuremath{\mathrm{#1}}}
\newcommand{\variable}[1]{\ensuremath{\bm{\mathrm{#1}}}}

\newcommand{\bx}{\bm{x}}

\newcommand{\bz}{\bm{z}}

\providecommand{\vector}[1]{}
\renewcommand{\vector}[1]{\mathmbox{\vec{#1}}}

\newcommand{\balpha}{\variable{\vector{\alpha}}}

\newcommand{\xsrc}{\vector{p}}
\newcommand{\bxsrc}{\vector{\variable{p}}}
\newcommand{\ximg}{\vector{\xi}}
\newcommand{\bximg}{\vector{\variable{\xi}}}

\newcommand{\by}{\variable{y}}
\newcommand{\flux}{f}
\newcommand{\bflux}{\variable{\flux}}
\newcommand{\T}{\varmatrix{T}}

\begin{document}

\maketitle

\begin{abstract}
    The analysis of optical images of galaxy-galaxy strong gravitational lensing systems can provide important information about the distribution of dark matter at small scales. However, the modeling and statistical analysis of these images is extraordinarily complex, bringing together source image and main lens reconstruction, hyper-parameter optimization, and the marginalization over small-scale structure realizations. We present here a new analysis pipeline that tackles these diverse challenges by bringing together many recent machine learning developments in one coherent approach, including variational inference, Gaussian processes, differentiable probabilistic programming, and neural likelihood-to-evidence ratio estimation.  Our pipeline enables: (a) fast reconstruction of the source image and lens mass distribution, (b) variational estimation of uncertainties, (c) efficient optimization of source regularization and other hyperparameters, and (d) marginalization over stochastic model components like the distribution of substructure. We present here preliminary results that demonstrate the validity of our approach.
\end{abstract}

\section{Introduction}

The existence of dark matter, its overall abundance and its distribution at galactic scales are reasonably well-understood thanks to its gravitational interactions. However, determining the properties of the fundamental constituents of dark matter remains one of the major unresolved issues in physics~\citep{Bertone:2004pz}.

Detection of the smallest dark matter structures on subgalactic scales is a potential inroad. In the standard $\Lambda$CDM model of cosmology, dark matter is cold and noninteracting. As a result, $n$-body simulations predict that galactic dark matter halos should contain exponentially-many lower-mass substructures called \emph{subhalos}. The subhalo mass function $dn/dm$ describing the number of subhalos at a given mass can be different if dark matter is instead warm or has nongravitational interactions. In these theories, overdensities below a particular threshold are unable to collapse into subhalos, and the subhalo mass function drops to zero below some threshold (see e.g. \citet{Viel:2013fqw,Schneider:2011yu}).

Since star formation is suppressed in subhalos with mass below $\sim 10^9\, M_\odot$, it is difficult to detect light from these halos directly. Instead, searching for their gravitational effects is a promising approach. In this work we study strong galaxy-galaxy gravitational lenses, where a source galaxy's light is dramatically distorted and multiply imaged by an intervening lens galaxy. The gravitational influence of subhalos imprints additional percent-level perturbations on the resulting observed image. To date, analyses of these systems have yielded two subhalo detections~\citep{Vegetti:2009cz,Vegetti:2012mc} 

Fitting a lensing system and properly marginalizing over variations in the source, lens and subhalos to constrain properties of substructure is an extremely difficult problem. This is due to the dramatic range of possible source galaxy light distributions and the large number of parameters that must be marginalized over in realistic lensing models in order to perform substructure inference. On one hand, methods such the one in \citet{Vegetti:2008eg} use Bayesian lensing models to closely fit lensing observations and conventional sampling methods to compute posteriors for subhalo parameters. However, conventional sampling methods require sampling from the joint posterior over all model parameters. These methods are thus restricted to searching for effects from one or two subhalos. \revised{On the other hand, since many works have shown neural networks have the capacity to measure main lens parameters~\citep{Hezaveh:2017sht,Morningstar:2019szx} and detect subhalos~\citep{DiazRivero:2019hxf,Ostdiek:2020cqz,Ostdiek:2020mvo}, \citet{Brehmer:2019jyt} leveraged them through likelihood-free inference to marginalize over large numbers of subhalos and directly compute marginal posteriors for subhalo mass function parameters.} However, this approach is extremely data-hungry when applied to even toy problems, since it requires amortizing over all possible variations in lensing systems.

In this work we merge the strengths of both approaches to perform targeted inference on individual lensing systems. In the first stage of our analysis, we introduce a new model for lensing systems that describes the source light distribution using an approximate Gaussian process~\citep{gp}. By implementing this in an automatically-differentiable manner and leveraging variational inference~\citep{DBLP:journals/corr/cs-AI-9603102,Jordan99anintroduction,JMLR:v14:hoffman13a,vi}, we fit approximate posteriors for all $\mathcal{O}(10^5)$ source and lens parameters to an observation using gradient descent. By sampling from this approximate posterior, we generate training data that looks \emph{very similar} to the observation. In the second analysis step, we use this data to train neural networks using the tools of likelihood-free inference to produce posterior distributions for substructure parameters. Applying these networks to the observation of interest yields inference results.

In the remainder of this paper, we describe our new model for lensing systems, our inference procedure, and present results from analyzing a mock lensing system. In upcoming work we will elaborate on our new source model~\citep{sourcemodeling} and substructure inference approach~\citep{substructureinference}.

\section{Modeling}

Here we present a flexible Bayesian lensing model that has the capacity to model complex, high-resolution images and can be fit using gradient descent. The base model has three components\revised{: a lens, with parameters denoted $\bz_l$, a source, parametrized by $\bz_s$, and the instrument response, which here is simply pixel noise assumed to be normally distributed with known standard deviation. Ultimately, we consider all of the parameters of the base model as nuisance parameters that we will need to marginalise out in order to infer an additional set of lens dark matter substructure parameters (e.g. the position and mass of a single subhalo, as we present below), denoted $\bz_d$.}

The lens models the presence of gravitating mass along the line of sight which bends the path of light (see \cref{fig:geometry}). In the thin lens formalism this process is described by the \emph{lens equation} \revised{$\bxsrc = \bximg - \balpha(\bximg)$}, which relates image plane coordinates \revised{$\bximg$} to source plane coordinates $\bxsrc$ via the displacement field $\balpha$. \revised{These are all arrays of $N$ 2-dimensional vectors, where $N$ is the number of pixels in the analysed image}. \revised{The displacement $\balpha$} is a linear function of the overall projected mass distribution $\Sigma(\ximg)$ in the image plane, which means that different mass components can be freely superposed. The displacement due to even a single component, however, can be a complex function of position, \revised{and in images of interest leads to the projection of disjoint sections of the image onto the same region in the source plane.}

We include two smooth analytic components for the lens: a singular power-law ellipsoid (SPLE) for the main lens galaxy and external shear (see e.g. \citet{Chianese:2019ifk} for the displacement fields). \revised{This model has previously been used to model observational data and has been shown capable of modeling the combined distribution of dark and baryonic matter in the inner regions of galaxies at the percent level \citep{Suyu:2008zp}}. It contains a total of eight parameters\revised{, which form $\bz_l$.}


We use a novel approach to modeling the sources in strongly lensed systems which has two main aims: to have a regularising effect so that the lens parameters can be constrained, and to treat source uncertainties so that they can be disentangled from the effect of lensing substructure. This is achieved with a generative model inspired by Gaussian processes, briefly described as:
\begin{equation}
    \bflux = \T(\bxsrc, \sigma)\, \by, \quad \text{with} \quad \by \sim \mathcal{N}(\variable{0}, \alpha^2).
\end{equation}
Here $\by$ is an array of $N$ flux parameters associated to the image pixels and assumed to have a Gaussian prior, while $\T$ is a \emph{transfer matrix}, defined below such that the true source fluxes $\bflux$ have a given covariance $\varmatrix{K} = \alpha^2 \T\T^T$. We use a Gaussian radial basis function to model the covariance between fluxes at two \emph{points} in the source plane:
\begin{equation}\label{eqn:covariance-function}
    \operatorname{cov}(\flux(\xsrc_1), \flux(\xsrc_2)) = k(\xsrc_1, \xsrc_2) = \alpha^2 \exp\qty(- \frac{\abs{\xsrc_1 - \xsrc_2}^2}{2 \sigma^2}).
\end{equation}
Its hyperparameters are $\alpha^2$ and $\sigma$, describing respectively the prior variance and the correlation scale in the source plane.

\revised{However, we are interested in the covariance matrix of the light received in \emph{pixels}, not at points. This depends on the intrinsic covariance between different points in the source (captured by \cref{eqn:covariance-function}) as well as the covariance due to the overlap between pixels projected back into the source plane. Accounting for this requires integrating \cref{eqn:covariance-function} over the overlap between each pair of back-projected pixels. Since this is extremely slow to compute exactly, we approximate each back-projected pixel using a} 2D Gaussian (see \cref{fig:pixels}): $g_i(\xsrc) = \mathcal{G}(\xsrc - \xsrc_i, \Sigma_i)$, with covariance matrix $\Sigma_i$ derived to match the shape and total area of the pixel. Then the covariance matrix becomes
\begin{equation}
    \operatorname{cov}(\bflux)_{ij} = \varmatrix{K}_{ij} = \iint k(\xsrc_1, \xsrc_2)\ g_i(\xsrc_1) g_j(\xsrc_2) \ \dd[2]{\xsrc_2}\dd[2]{\xsrc_2}
    = 2\pi \alpha^2 \sigma^2 \mathcal{G}(\xsrc_i - \xsrc_j, \Sigma_i + \sigma^2 I + \Sigma_j).
\end{equation}

\revised{Usually, one needs to calculate the transfer matrix $T$ from $K$ by e.g. matrix square root or Cholesky decomposition, but due to the high dimensionality of the problem these operations are infeasible. Instead, we realise that the outer product $\T\T^T$ is akin to a spatial convolution, and that the convolution of a Gaussian with itself is a Gaussian with twice the variance, which allows us to approximate $\T_{ik} \sim \mathcal{G}(\xsrc_i - \xsrc_k, \Sigma_i + \frac{\sigma^2}{2} I)$. The proper normalisation depends on the number density of pixels in the source plane; we give a full derivation in \citet{sourcemodeling}.}

Finally, instead of allowing the kernel size to vary, we consider a number of independent \emph{GP layers} with fixed $\sigma_{(k)}$ ranging from galaxy- to pixel-size in order to model details on various scales. Each layer thus has an independently-optimised array of flux parameters $\by_{(k)}$ and overall variance $\alpha_{(k)}^2$. The collection of all $\by_{(k)}$ forms the source parameter array $\bz_s$, which together with the hyperparameters $\alpha_{(k)}^2$ are optimized variationally.

\revised{We implement our lensing model using the \texttt{pytorch} and \texttt{pykeops} libraries, which ensures the model output is automatically-differentiable with respect to all (hyper)parameters: $\bz_l$, $\bz_s$, $\alpha_{(k)}^2$, and enables us to perform all calculations on graphical processing units.}

\section{Statistical analysis}

The overall analysis strategy splits in two steps.
\begin{enumerate}
    \item Fit an approximate posterior $q_\phi(\bz_s, \bz_l|\bx_0)$ for the lens and source parameters $\bz_l$ and $\bz_s$ to an observation $\bx_0$ using variational inference, simultaneously optimizing the GP hyperparameters $\theta = \{ \alpha_{({k})}^2 \}$.
    \item Sample data from this constrained model to train an inference network to predict $p(\bz_d|\bx_0)$, the marginal posterior for substructure parameters $\bz_d$ for the observation.
\end{enumerate}

\paragraph{Variational inference}

We approximate the posterior over the source and lens parameters with a multivariate normal distribution over the lens parameters and a diagonal normal distribution over the source parameters:
\begin{equation}
    q_\phi(\bz_l, \bz_s|\bx_0) = \mathcal{G}_{\phi_l}(\bz_l)\, \mathcal{N}_{\phi_s}(\bz_s),
\end{equation}
where $\phi = \{ \phi_l, \phi_s \}$ denote the distribution's mean and covariance parameters. This approximation is justified since all parameters are reasonably well-constrained. While it neglects source parameter correlations, the diagonal normal guide for the $\mathcal{O}(10^5)$ source parameters is necessary since the matrix inversion required for standard GP inference~\citep{gp} is prohibitively expensive.  \revised{Since close-by source parameters are in general anti-correlated, we find that the neglect of correlations increases the variance of the posterior predictive distribution, which is consistent with the goal of using samples from the fitted model for training targeted neural inference networks. We leave a quantitative study of these effects to future work.}

In order to optimize the above approximate posterior with respect to $\phi$ and also optimize the GP hyperparameters, we use gradient descent to maximize the evidence lower bound (ELBO)~\citep{DBLP:journals/corr/cs-AI-9603102,Jordan99anintroduction,vi,JMLR:v14:hoffman13a} for the observation, $\mathbb{E}_{\bz \sim q_\phi(\bz|\bx_0)} \left[ \ln p_\theta (\bx_0, \bz) - \ln q_\phi (\bz|\bx_0) \right]$.  \revised{To this end, we use the probabilistic programming language \texttt{pyro}, and the auto-differentiation capabilities of \texttt{pytorch}.}

\paragraph{Likelihood-free inference}

Next we seek the posterior $p(\bz_d | \bx_0)$ for the substructure parameters, marginalized over the lens and source parameters $\eta \equiv \{ \bz_l, \bz_s \}$. The integral over $\eta$ is intractible since it is very high-dimensional and the integrand is a complex nonlinear function. Instead, we approximate $p(\bz_d | \bx_0)$ using neural likelihood-to-evidence ratio estimation techniques~\citep{Hermans:2019ioj}.

The strategy is to estimate the ratio $r(\bx, \bz_d) \equiv p(\bz | \bx) / p(\bz)$, from which the marginal posterior is easily recovered. To do this we train a classification network $d(\bx, \bz_d) \in [0, 1]$ to discriminate the hypotheses. In the first, the data and substructure parameters are drawn jointly from the parameter priors and model: $\bx, \bz_d \sim p(\bx, \bz_d)$. In the second, the data and substructure parameters are sampled marginally: $\bx, \bz_d \sim p(\bx) p(\bz_d)$. We adopt the binary cross-entropy loss function and optimize the parameters of the classification network using gradient descent. After this we can compute $r(\bx_0, \bz_d) = d(\bx_0, \bz_d) / (1 - d(\bx_0, \bz_d))$ and obtain the substructure parameter posteriors for the observation of interest.

Formally, marginalizing over the source and lens requires training the classification network on all observationally possible lensing images. Framing this in terms of our lensing model, generating this training dataset requires sampling the lens, source and substructure parameters from their priors, computing the lensed image $\mu$, and generating an observation by adding Gaussian pixel noise with standard deviation $\sigma_n$:
\begin{equation}
    \bx \sim \mathcal{N}(\bx | \mu(\bz_l, \bz_s, \bz_d), \sigma_n) \, p(\bz_l) \, p(\bz_s) \, p(\bz_d).
\end{equation}
Since we eventually are interested in sub-percent variations in the images, this is an extraordinarily complex endeavor, which would require a very large amount of training data as well as a network with very high capacity.

Instead, we propose a much simpler and more tractable approach where we \emph{target} the analysis on a particular observation by constraining the sampling of the lens and source parameters. In particular, we sample $\bz_l, \bz_s \sim q_\phi(\bz_l, \bz_s | \bx_0)$ rather than from their priors. This training data generation procedure excludes lensing systems incompatible with the observation being analyzed, dramatically decreasing the amount of data required for the classification network, allowing us to use simple architectures and shorter training times.

\section{Results and discussion}

\begin{figure}[b]
    \centering
    \includegraphics[width=\linewidth]{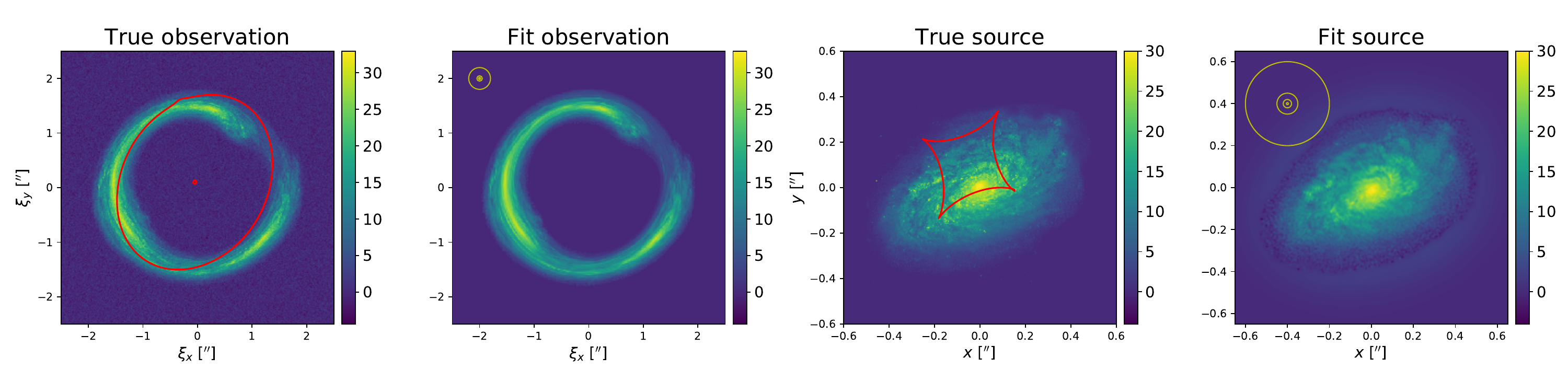}
    \caption{From left to right: the mock observation, fit observation, true source and fit source. The displayed fits are computed using the mean estimates of the source and lens parameters from the approximate posterior $q_\phi(\bz | \bx)$. The red curves show the critical curve and caustic of the main lens. The yellow circles indicate the spatial scales of the five Gaussian process layers, the smallest of which is not visible here.}
    \label{fig:elt-mock-fit}
\end{figure}

\begin{figure}[b]
    \centering
    \includegraphics[width=0.8\linewidth]{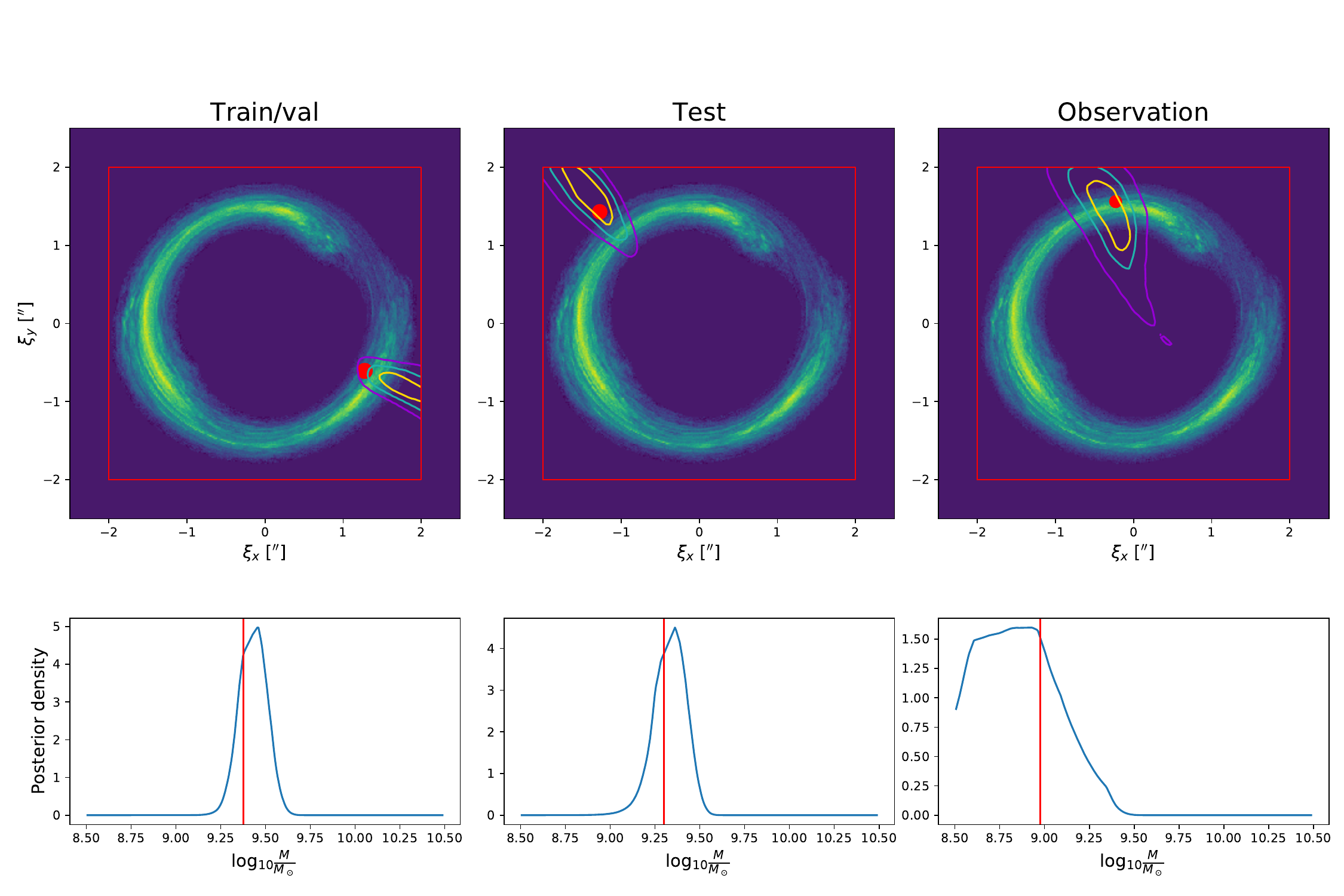}
    \caption{Preliminary results from our pipeline: posteriors for the subhalo position and mass for a systems from the training/validation set (left panel), a system from a held-out test set (middle), and the observation being analyzed (right). The purple, teal and yellow contours in the first row show the $99.7\%$, $95\%$ and $68\%$ containment regions. The red dots indicate the true subhalo position and the red box is the area from which the subhalos were uniformly sampled. The red lines in the second row mark the true subhalo masses. These were obtained with only 10,000 training examples.}
    \label{fig:sh-posteriors}
\end{figure}

As a preliminary test of our analysis procedure, we construct a $400\times 400$ mock observation based on the Extremely Large Telescope's~\citep{elt} capabilities that contains a single $\sim10^9\, M_\odot$ subhalo. The mock observation and source are shown in \cref{fig:elt-mock-fit}, along with the mean observation and source reconstructed with a five-layer GP source model in our first analysis step. The observation reconstruction closely matches the mock observation, and captures very fine details in the source. The inferred values of the lens parameters are within $\sim 0.1 - 5\%$ of the true parameter values. For the second analysis step, we train simple neural networks built from convolutional and fully-connected layers to give 2D and 1D posteriors for the subhalo's position and mass, marginalized over the 174,458 source and lens parameters. The resulting posteriors, obtained with just 10,000 training examples, are shown in \cref{fig:sh-posteriors}. Our subhalo position reconstructions are accurate, with the true position of the subhalo in the mock observation (right column) lying within the $68\%$ containment region. The mass posterior is centered close to the true subhalo mass, though it is reasonably broad. The whole analysis runs in a few hours on a single Nvidia GTX TitanX GPU.

Applying our pipeline to real data will require extending the lensing model. A more realistic version of our lensing system would contain $\sim 320$ subhalos above $10^5\, M_\odot$ within the observation region. Dark matter halos along the line of sight are additionally expected to contribute $3-10$ times more lensing distortions than subhalos~\citep{Despali:2017ksx}. Lastly, we have omitted the main lens' light. All of these components are straightforward to incorporate in our framework. In upcoming work  we will apply this extended analysis to search for subhalos and set constraints on the subhalo mass function using existing lensing observations in future work. More generally, we anticipate our approximate Gaussian process model and targeted likelihood-free inference strategy will find other exciting applications to difficult astrophysical imaging and data analysis problems.


\section*{Broader Impact}

This work is focusing on the precision analysis of astronomical images based on forward models. Variants of the presented approach could be applicable to other areas of the physical sciences.  Although we do not anticipate potential for misuse of the presented methods or the danger of grossly biased results, the usual care has to be exercised when drawing scientific conclusions based on a complex analysis machinery.

\begin{ack}
We thank Marco Chianese, Camila Correa, Gilles Louppe, Ben Miller and Simona Vegetti for useful discussions.

This work uses \texttt{numpy}~\citep{Harris_2020}, \texttt{scipy}~\citep{2020SciPy-NMeth}, \texttt{matplotlib}~\citep{Hunter:2007}, \texttt{pyro}~\citep{bingham2018pyro}, \texttt{pytorch}~\citep{NEURIPS2019_9015}, \texttt{pykeops}~\citep{charlier2020kernel}, \texttt{jupyter}~\citep{Kluyver:2016aa} and \texttt{tqdm}~\citep{casper_da_costa_luis_2020_4054194}. Computations were carried out on the DAS-5~\citep{das5} cluster, which is funded by the Netherlands Organization for Scientific Research (NWO/NCF).
\end{ack}

\small

\bibliography{refs}

\clearpage

\appendix
\section*{Appendix}

\vfill

\begin{figure}[h!]
    \centering
    \includegraphics[width=0.6\textwidth]{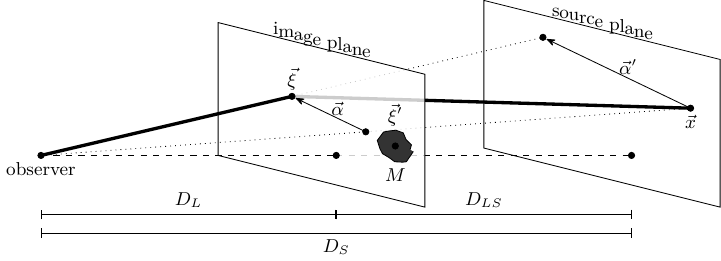}
    \caption{Geometry of strong lensing systems. The lensing mass $M$ located at $\vec{\xi}'$ bends the light ray (thick line) emanating from the point $\vec{x}$, so that to the observer it looks like it is coming from the direction of $\vec{\xi}$. General relativity predicts the deflection angle $\vec{\alpha}'$ as viewed from the image plane based on the mass $M$ and the distance $\vec{\xi} - \vec{\xi}'$. It then has to be rescaled by $D_{LS} / D_S$ to obtain the displacement field $\vec{\alpha}$ (as viewed by the observer). Angles are all assumed to be small enough that they can be used for Euclidean calculations. The dashed line is the optical axis perpendicular to the planes and connects the origins of the coordinate systems for each plane.}
    \label{fig:geometry}
\end{figure}

\vfill

\begin{figure}[h!]
    \centering
    \includegraphics[width=0.2\textwidth]{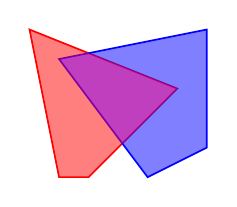}
    \includegraphics[width=0.2\textwidth]{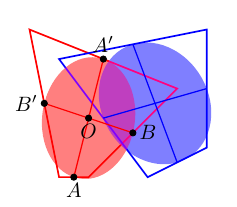}
    \includegraphics[width=0.2\textwidth]{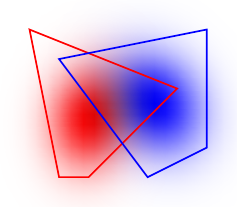}
    \caption{Our Gaussian approximation of pixels projected into the source plane. The projection of two pixels onto the source plane (a). A full treatment should consider their exact shapes and use an indicator function that is nonzero only inside the shaded areas. Instead, the projections are approximated to ellipses defined by the projections of their centers and the vectors ${A A'}$ and ${B B'}$ (b). The indicator functions are Gaussians with covariances derived from those ellipses (c).}
    \label{fig:pixels}
\end{figure}

\vfill

\end{document}